\documentclass[useAMS,usenatbib,usegraphicx,onecolumn]{biom}

\makeatletter
  \renewcommand\normalsize{\@setfontsize\normalsize\@xiipt{14.5}% From size12.clo
   \abovedisplayskip 4pt plus 1pt minus 1pt
   \belowdisplayskip \abovedisplayskip
   \abovedisplayshortskip 4pt plus 1pt
   \belowdisplayshortskip \abovedisplayshortskip
   \let\@listi\@listI}
  \def\@evenfoot{\centerline{\thepage}}
  \def\@oddfoot{\centerline{\thepage}}
  \def\@evenhead{}
  \def\@oddhead{}
  \gdef\@journal{\fbox{\vbox{
      This is the peer reviewed version of the following article:
      Meyer, S., Elias, J. and H\"ohle, M. (2012).
      A Space-Time Conditional Intensity Model for Invasive Meningococcal Disease Occurrence.
      \textit{Biometrics} \textbf{68}, 607--616,
      which has been published in final form at
      \url{http://dx.doi.org/10.1111/j.1541-0420.2011.01684.x}.
      This article may be used for non-commercial purposes in accordance with
      Wiley Terms and Conditions for Self-Archiving.
    }}}
\makeatother

\usepackage[latin1]{inputenc}  % character encoding of LaTeX source files
\usepackage[T1]{fontenc}       % T1-encoding for european languages with a latin alphabet
\usepackage[USenglish]{babel}  % language of the manuscript is American English
\usepackage{graphicx}

\usepackage{lmodern}           % Replacement for Computer Modern fonts; alternative: Palatino (or mathpazo if text contains math)
\usepackage{amsmath, amsfonts, amssymb}
\usepackage{bm}                % \bm for bold mathematical symbols, this is better than \boldsymbol from amsmath
\usepackage{bbm}               % \mathbbm for double-stem mathematical symbols like R, N or 1 (indicator)
\usepackage{subfigure}
\usepackage{url}               % for formatting URLs, e.g. in the bibliography (manages line breaks in URLs)
\usepackage{natbib}            % adds citation styles
\usepackage{Sweave}

 \newcommand{\comment}[1]{}
 \newcommand{\TODO}[1]{}
\newcommand{\dif}{\,\mathrm{d}}
\newcommand{\R}{\mathbbm{R}}
\newcommand{\N}{\mathbbm{N}}
\renewcommand{\P}{\mathbbm{P}}

\newcommand{\id}{\mathbbm{1}}
\newcommand{\abs}[1]{\lvert#1\rvert}
\newcommand{\norm}[1]{\lVert#1\rVert}
\newcommand{\offset}{\rho_{\tau(t),\xi(\bm{s})}}

\usepackage[outerbars,xcolor]{changebar}
\ifx\pdfoutput\undefined
\else\ifnum\pdfoutput>0
  \usepackage{pdfcolmk}
\fi\fi
\cbcolor{red}

\newcommand{\dckeywords}{Conditional intensity function; Infectious disease surveillance data; Spatio-temporal point process; Stochastic epidemic modelling}

\title[A space-time conditional intensity model for IMD occurrence]{\LARGE A space-time conditional intensity model\\for invasive meningococcal disease occurrence}
\author{Sebastian Meyer$^{1,2,*}$\email{Sebastian.Meyer@med.uni-muenchen.de},
Johannes Elias$^{3}$, and
Michael H\"ohle$^{4,2,**}$\email{HoehleM@rki.de} \\
$^{1}$Department of Psychiatry and Psychotherapy, Ludwig-Maximilians-Universit\"at, M\"unchen, Germany \\
$^{2}$Department of Statistics, Ludwig-Maximilians-Universit\"at, M\"unchen, Germany \\
$^{3}$German Reference Centre for Meningococci, University of W\"urzburg, W\"urzburg, Germany \\
$^{4}$Department for Infectious Disease Epidemiology, Robert Koch Institute, Berlin, Germany}
\date{{\it Received November} 2010. {\it Revised August} 2011.  {\it Accepted August} 2011.}
\pagerange{\pageref{firstpage}--\pageref{lastpage}}
\volume{00}
\pubyear{2011}
\artmonth{October}
\doi{10.1111/j.1541-0420.2011.01684.x}

\begin{document}
\label{firstpage}

%-ABSTRACT--------------------------------------------------------

\begin{abstract} %\noindent \textsc{Summary}.
  A novel point process
  model continuous in space-time is proposed for quantifying the
  transmission dynamics of the two most common meningococcal antigenic
  sequence types observed in Germany 2002--2008. Modelling is based on
  the conditional intensity function (CIF) which is described by a
  superposition of additive and multiplicative components.  As an
  epidemiological interesting finding, spread behaviour was shown to
  depend on type in addition to age: basic reproduction numbers were
  0.25 (95\% CI 0.19-0.34) and 0.11 (95\% CI 0.07-0.17) for types
  B:P1.7-2,4:F1-5 and C:P1.5,2:F3-3, respectively. Altogether, the
  proposed methodology represents a comprehensive and universal
  regression framework for the modelling, simulation and inference of
  self-exciting spatio-temporal point processes based on the
  CIF. Usability of the modelling in biometric practice is promoted by
  an implementation in the \texttt{R} package \texttt{surveillance}.
% \begin{flushleft}
% \textsc{Key words}: \dckeywords.
% \end{flushleft}
\end{abstract}


%-TITLE PAGE------------------------------------------------------

\maketitle

%-MAIN PART-------------------------------------------------------

\section{Introduction}
\label{sec:intro}

Infectious diseases -- such as influenza, gastroenteritis, and the
``swine flu'' among humans, or foot and mouth disease, the ``bird
flu'', and classical swine fever among animals -- are a matter of
tremendous public concern especially gaining attention in case of
outbreaks.
%@biometrics
%Collaboration of public health decision makers,
%veterinaries, microbiologists, epidemiologists, statisticians and many
%others is indispensable for understanding and controlling disease
%dynamics.
%The statistician's contribution is typically based on
%stochastic epidemic models inheriting from the stochastic
%susceptible-infectious-recovered (SIR) model described by, e.g.,
%\citet{AnderssonBritton2000}.
The present work concentrates on stochastic modelling and associated
inference for spatio-temporal epidemic point referenced data motivated
by the analysis of routinely collected invasive meningococcal disease
(IMD) data. IMD is a life-threatening human bacterial disease mostly
manifesting as meningitis or sepsis. Its pathogenic agent,
\emph{Neisseria meningitidis} (aka \emph{meningococcus}), can be
transmitted by large droplet secretions from the respiratory tract of
colonized or infected humans. The only reservoir of meningococci is
the human (mostly nasopharyngeal) mucosa~\citep{IMDreview}.
%The incubation period is around four days, but it actually exhibits
%large deviations between different cases ranging from one to ten days
%\citep[p.\,415--421]{CCDM2008}.
Data on cases of IMD related to the two most common meningococcal
finetypes \mbox{B:P1.7-2,4:F1-5} and \mbox{C:P1.5,2:F3-3} in Germany
2002--2008 are obtained from the German Reference Centre for
Meningococci (Nationales Referenzzentrum f{\"u}r Meningokokken, NRZM).
Here, a 'finetype' represents a unique
combination of serogroup, sequence type of variable region 1 and 2 of
the outer membrane protein PorA, and sequence type of the variable
region of the outer membrane protein FetA. One specific question of
interest for the researchers at the NRZM is whether the two finetypes
(in what follows abbreviated B and C) exhibit different
spatio-temporal behaviour.

The postal code of the patient's home address was the spatial
resolution available for our analysis. Despite being spatially
discrete we consider centroids of postal code areas as
quasi-continuous in space when looking at entire Germany. As usual
with infectious diseases, the actual time point of infection is
unknown for the IMD cases. Therefore, we define the beginning of
illness and infectivity as the date of specimen sampling.

\setkeys{Gin}{width=0.475\textwidth}
\begin{figure}%[bht]
\centering
\subfigure[Finetype B:P1.7-2,4:F1-5.]{
\includegraphics{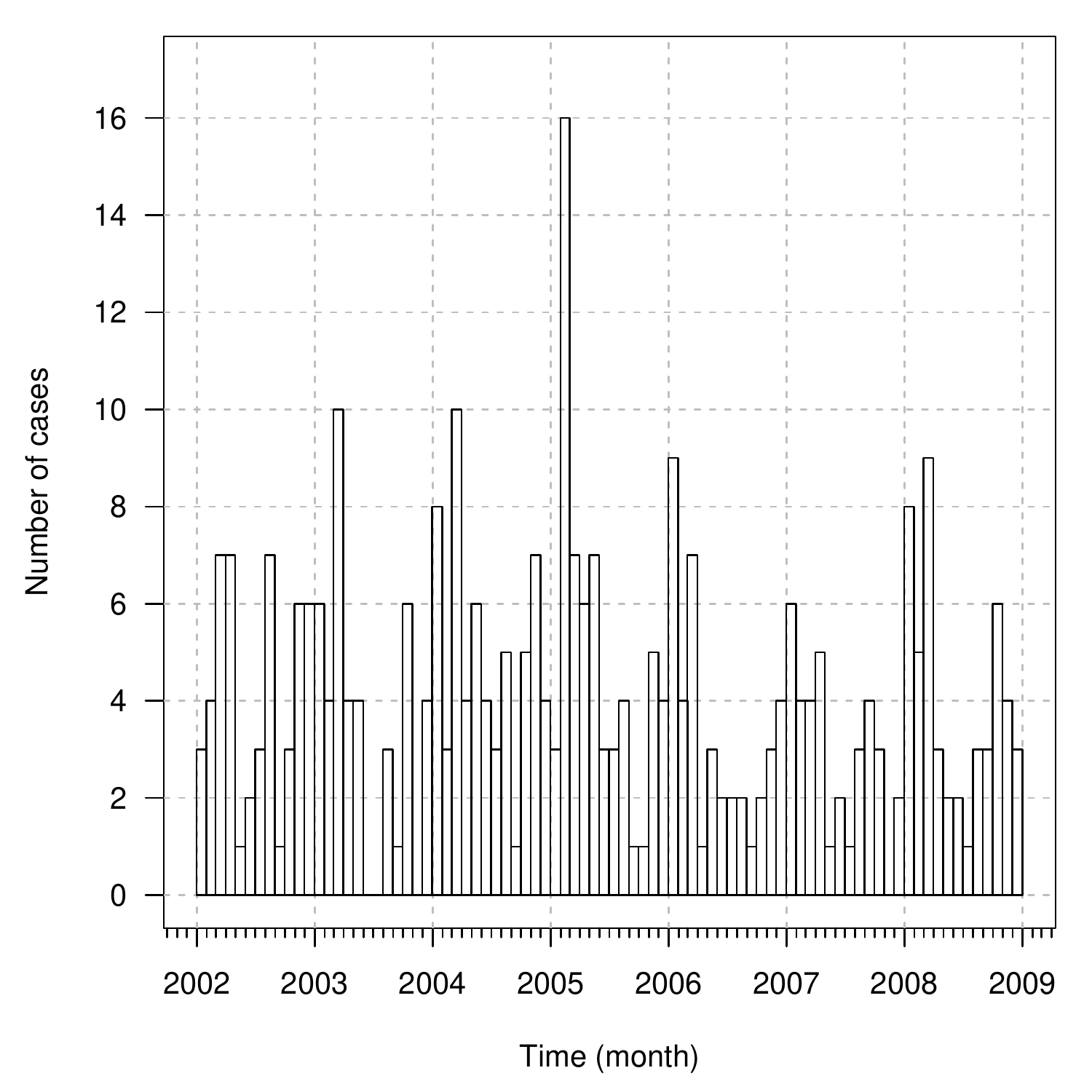}
\label{fig:IMD:temporal:B}
}
\subfigure[Finetype C:P1.5,2:F3-3.]{
\includegraphics{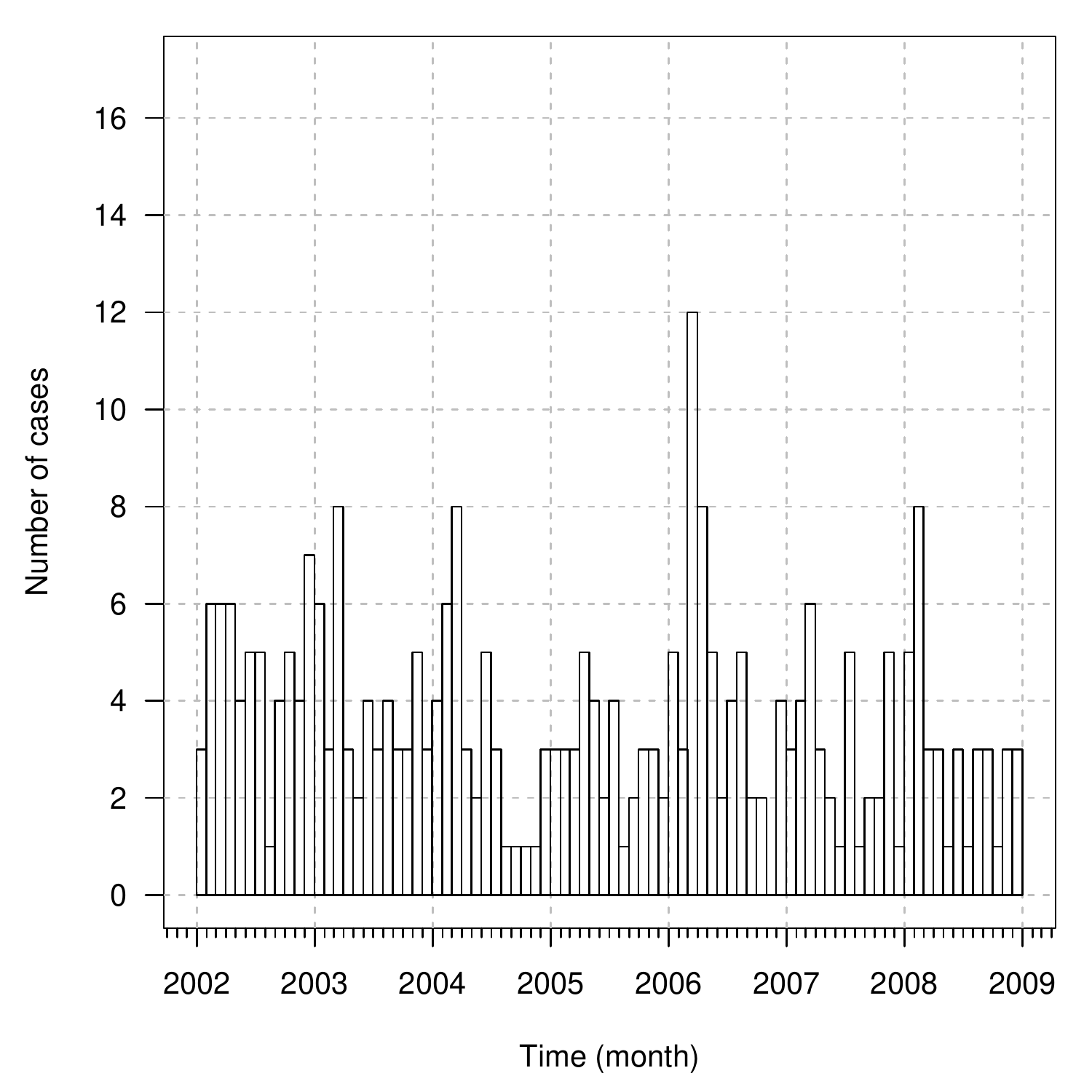}
\label{fig:IMD:temporal:C}
}
\caption{Monthly numbers of IMD cases for both finetypes separately.}
\label{fig:IMD:temporal}
\end{figure}
All in all, $n=636$ infections with finetypes
B (336) and
C (300) have been registered.
Figure~\ref{fig:IMD:temporal} shows the monthly numbers of IMD cases
for each finetype. Cases of IMD predominantly occur during winter and
early spring, which can be seen from more or less pronounced peaks in
the figure.
Specifically, a connection between outbreaks of meningococci
and influenza is hypothesized. For example, \citet{Jensen2004} found
an association between the influenza detection rate and the number of
IMD cases during the same week in temporal analysis of data from
Northern Jutland County in Denmark, during 1980--1999.

\setkeys{Gin}{width=0.475\textwidth}
\begin{figure}%[bht]
\centering
\subfigure[Finetype B:P1.7-2,4:F1-5.]{
\includegraphics{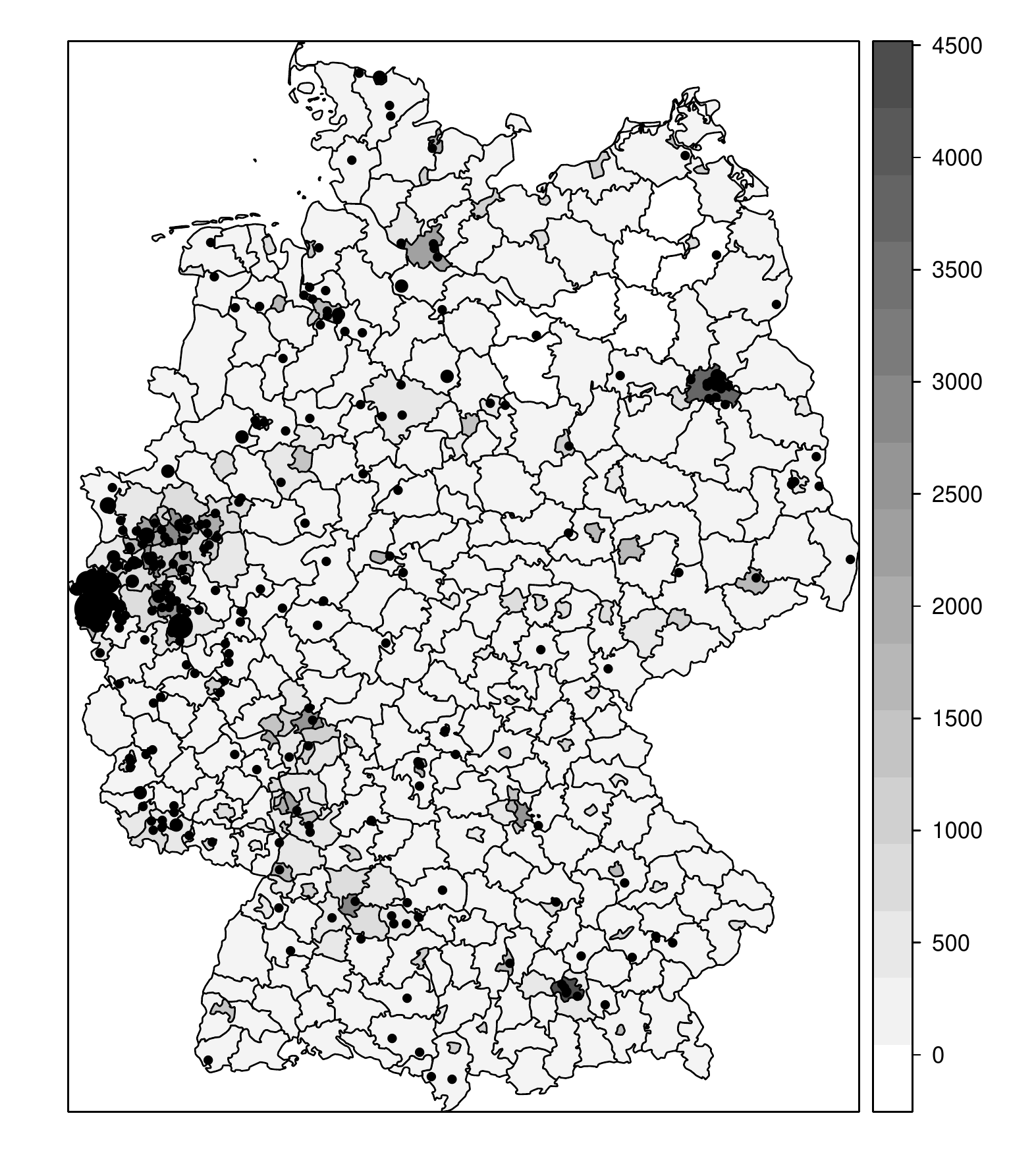}
\label{fig:IMD:spatial:B}
}
\subfigure[Finetype C:P1.5,2:F3-3.]{
\includegraphics{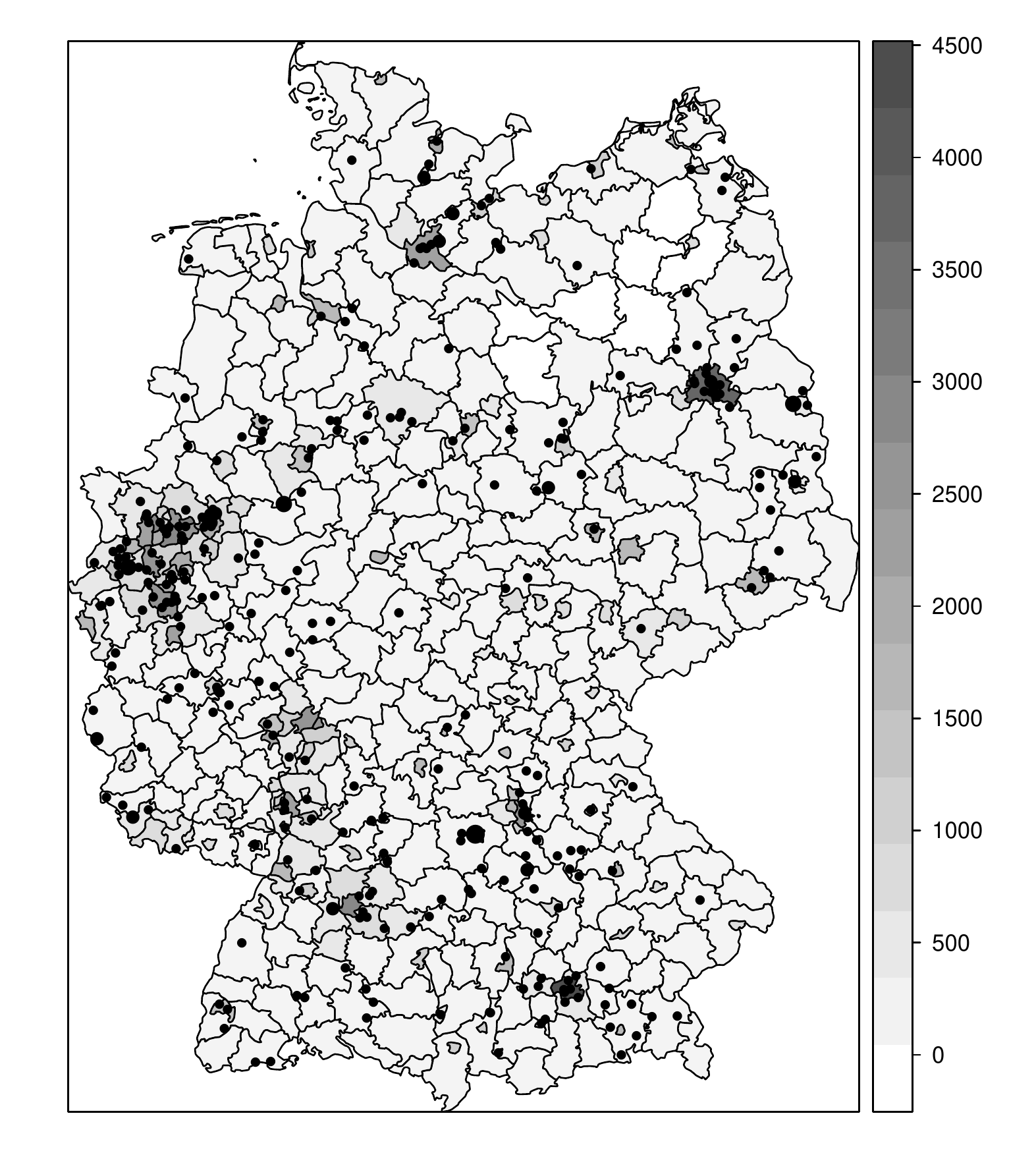}
\label{fig:IMD:spatial:C}
}
\caption{Spatial point patterns of the cases of meningococci by
  finetype during the years 2002--2008. The area of each dot is
  proportional to the number of cases at its location. Also shown are
  the population densities (inhabitants per km$^2$) of Germany's
  districts (source: \citet{DESTATIS}).}
\label{fig:IMD:spatial}
\end{figure}

Figure~\ref{fig:IMD:spatial} presents the spatial distributions of the
two finetypes based on the postcodes of the patients' residences. Over
the 7-year period some cases shared the same postal code, therefore,
the area of each point in the figure is drawn proportional to the
number of cases at its location. For the serogroup B finetype in
\subref{fig:IMD:spatial:B} the highest point multiplicity is $16$,
whereas for the serogroup C finetype in \subref{fig:IMD:spatial:C}
this number is $4$. In connection with the temporal occurrence of the
events shown in Figure~\ref{fig:IMD:temporal}, the spatial
distribution suggests that IMD is an endemic disease, i.e.\ cases can
occur at any time and at any location. The maps also show the
population densities of the districts, which can be assumed to be
roughly proportional to the population at risk of infection. Spatial
heterogeneity of the observed point patterns thus partially arises
from spatial variation in the population density.  Not surprisingly,
the intensity of points in metropolitan areas like Berlin, Munich or
the Ruhr is higher.  Animated graphics of the space-time locations of
infections give more insight into the epidemic character of the
finetypes, and can be found as Web Animation 1. Here, it appears as if
finetype B exhibits a more stationary pattern than finetype C -- in
the sense that infections cluster more in space and time. It is
supposed, yet not proven, that this phenomenon is due to differences
in the mucosal immune reaction elicited; specifically, finetype B
might be more successful than C in evading mucosal clearance.

Quantifying the dynamics of IMD would be an important step in the
finetype characterisation of IMD. We want to perform such an
investigation in a spatio-temporal manner and therefore use
spatio-temporal point processes as modelling framework. Specifically,
we want to establish a regression framework allowing us to quantify
the transmission dynamics of IMD and its dependency on
covariates. Point process modelling has in the context of epidemics
been used in a discrete spatial setting in, e.g.,
%@short@ \citet{LawsonLeimich2000},
\citet{neal2004}, \citet{Diggle2006}, \citet{Scheel2007} and
\citet{jewell_etal2009}.  Spatio-temporal epidemic modelling in an
explicit continuous spatial setting, however, is rare with
\citet{Diggle2005} being one of the few examples of covariate adjusted
modelling. One explanation is the balancing between optimal spatial
resolution of the data and confidentiality of cases.

Recently, there have been suggestions for splitting the dynamics of
infectious diseases into endemic and epidemic components; see
\citet{held_etal2005} for a discrete spatial -- discrete time
perspective and \citet{Hoehle2009} for a discrete spatial --
continuous time perspective. For the continuous spatial -- continuous
time setting, similar modelling approaches have been seen in the
analysis of earthquake data, see e.g.\ \citet{Ogata1998,Ogata1999}.
%Earthquakes
%and cases of infectious diseases have in common that they feature
%\emph{self-excitement}, i.e.\ events promote the future evolution of
%the point process by producing ``offspring'' events.
%Inspired by the
%self-exciting Hawkes process~\citep{Hawkes1971},
%\citet{Ogata1988} proposed the so-called \emph{epidemic-type
%  aftershock-sequences} (ETAS) model for earthquake occurrences which
%has also been extended to the spatio-temporal
%context~\citet{Ogata1998}.
Other areas of application
%drawing on similar modelling approaches
are the modelling of forest
fires~\citep{peng_etal2005}, residential
burglaries~\citep{mohler_etal2010}, and the analysis of bird nesting
patterns~\citep{Diggle2009}.
%In these applications modelling is,
%however, often tailored towards the specific use and thus there is a
%need for a more unifying regression approach for the modelling of
%space-time phenomena. Therefore, the additive-multiplicative CIF
%modelling is extended to the context of continuous space and process
%marks as motivated by the IMD application.
Altogether, our proposed modelling provides a unifying regression
framework -- beyond epidemics -- for the modelling, inference and
simulation of spatio-temporal point processes.

This article is organized as follows: Section~\ref{sec:stCIFmodel}
presents the spatio-temporal two-component epidemic model based on the
CIF, whereas Sections~\ref{sec:inference} and~\ref{sec:simulation}
discuss inference and simulation for the proposed
model. Section~\ref{sec:application} analyses the IMD data, and a
discussion in Section~\ref{sec:discussion} finalizes the article.

\section{Spatio-Temporal Two-Component CIF Model}
\label{sec:stCIFmodel} \label{sec:twinstim}

In the following text, we propose a novel additive-multiplicative model for the
conditional intensity function of an infectious disease process
continuous in space-time with events occuring in a prespecified
observation period $[0,T]$, $T>0$, and observation region $W\subset
\mathbbm{R}^2$. The CIF $\lambda^*(t,\bm{s})$ represents the instantaneous rate or hazard
for events at time $t$ and location $\bm{s}$ given all the
observations up to time $t$ (the asterisk notation shall represent the conditioning on the
random past history of the process).

The basic framework of the proposed model is to superimpose endemic
and epidemic components in order to model the IMD surveillance data
-- an idea similar to the two-component spatial SIR model~\citep{Hoehle2009}:
\begin{equation*} %\label{eq:twinstim-h+e}
	\lambda^*(t,\bm{s}) = h(t,\bm{s}) + e^*(t,\bm{s}) \qquad(t>0,
        \bm{s} \in W)\:.
\end{equation*}
The epidemic component $e^*(t,\bm{s})$ represents the spread of the
disease by person-to-person contact. The endemic component
$h(t,\bm{s})$ models otherwise imported cases and is -- contrary to
the epidemic component -- independent of the internal history of the
process.

\subsection[Specification of the Endemic Component]{Specification of the Endemic Component $h(t,s)$}

The endemic component is of the multiplicative form $h(t,\bm{s}) =
\rho(t,\bm{s}) \exp(\bm{\beta}'\bm{z}(t,\bm{s})),$ where
$\rho(t,\bm{s})$ is a known spatio-temporal intensity offset, e.g.\
the population density at time $t$ in the district containing the
location $\bm{s}$, such that the endemic rate of infection is
proportional to the population density. Furthermore,
$\bm{z}(t,\bm{s})$ is a linear predictor of endemic covariates, e.g.,
this could be a temporal trend or exogenous covariates resulting from
another jointly evolving point process. For example, in the IMD
application, an endemic covariate is the number of influenza cases on
a week\,$\times$\,district grid (possibly time-lagged).  Altogether,
the endemic component is modelled as a piecewise constant function on
some spatio-temporal grid resulting from a decomposition of the time
period $(0,T]$ and the observation region $W$. The consecutive time
intervals of this decomposition (e.g.\ weeks) are denoted by
$C_1,\ldots,C_D \subset (0,T]$, and the spatial tiles (e.g.\
districts) are denoted by $A_1,\ldots,A_M \subset W$. Let the
functions $\tau(t)$ and $\xi(\bm{s})$ return the indices of the
temporal and spatial grid units containing time point $t$ and
coordinate $\bm{s}$, respectively. Then, the endemic component can be
written as
\begin{equation} \label{eq:twinstim-h-grid}
	h(t,\bm{s}) = \offset \,\exp\big(\bm{\beta}'\bm{z}_{\tau(t),\xi(\bm{s})} \big)\:,
\end{equation}
where $\offset$ is the known interval- and tile-specific offset and $\{
\bm{z}_{\tau,\xi}: \tau \in \{1,\ldots,D\},\,\xi\in\{1,\ldots,M\} \}$
is a collection of covariates on the spatio-temporal grid
$\{C_1,\ldots,C_D\}\times\{A_1,\ldots,A_M\}$.

\subsection[Specification of the Epidemic Component]{Specification of the Epidemic Component $e^*(t,s)$}

The self-exciting component of the model essentially provides a
description of the infection pressure at a space-time location
$(t,\bm{s})$ caused by each infectious individual. This infectivity of
an infectious individual $j$, denoted by $e_j(t,\bm{s})$, corresponds
to the inhomogeneous rate of a Poisson process, the realisations of
which are the space-time locations of infected individuals. This so called
triggering function is factorized into separate effects of marks,
elapsed time, and relative location:
\begin{equation} \label{eq:twinstim-ej}
	e_j(t,\bm{s}) = e^{\eta_j} \, g(t-t_j) \,
                         f(\bm{s}-\bm{s}_j)\:, \qquad (t>t_j)
\end{equation}
where
$(t_j,\bm{s}_j)$ is the infection time and location of individual j,
$\eta_j = \gamma_0 + \bm{\gamma}'\bm{m}_j$ is a linear predictor
based on the vector of unpredictable marks $\bm{m}_j$ attached to the
infected individual, and $g$ and $f$ are positive temporal and spatial
interaction functions, respectively. The effects $\bm{\gamma}$ of
marks reflect that different individuals might cause more or less
secondary cases, depending on individual characteristics.

The interaction functions describe the decay of infectivity with an
increasing spatial or temporal distance from the infection
source.
% Note that an additive decomposition in \eqref{eq:twinstim-ej}
% would not have been reasonable: the distance of previously infected
% individuals $j$ to a location $s$ would then contribute to the total
% infection pressure at this location even if $g(t-t_j)\approx 0$,
% i.e.\ when the individual $j$ actually is no longer infective. In
% contrast, the multiplicative interaction of the temporal and spatial
% components in \eqref{eq:twinstim-ej} arranges for the desired zero
% contribution of individuals with $g(t-t_j)=0$.
In infectious disease applications, $f$ is often taken to be a
radially symmetric kernel corresponding to an isotropic spread of the
disease, such that \mbox{$f(\bm{s}-\bm{s}_j) \equiv
  f(\norm{\bm{s}-\bm{s}_j})$}. A typical example is
%biometrics shortening
%\begin{equation} \label{eq:f-gaussian}
%	f(\bm{s}) = \exp\left(-\frac{\norm{\bm{s}}^2}{2\sigma^2}\right) %\qquad (\bm{s}\in\R^2,\,\sigma>0) \:,
%\end{equation}
%i.e.\
to let $f$ be the kernel of a bivariate normal density with
zero mean and diagonal covariance matrix.
The temporal interaction function could be chosen as
%@short%
%\begin{equation} \label{eq:g-expdecay}
%	g(t) = e^{-\alpha t} \qquad (t>0,\,\alpha>0)
%\end{equation}
$g(t) = e^{-\alpha t}$, $t>0$, $\alpha>0$, representing an exponential
temporal decay of infectivity \citep{Hawkes1971}.
%shorten for biometric practice
%If $f$ or $g$ are modelled as constant
%functions equal to 1, individuals spread the disease homogeneously in
%space or time, respectively.

The resulting epidemic component $e^*(t,\bm{s})$ is the sum of
the contributions \eqref{eq:twinstim-ej} of all infectious individuals
at time $t$ and location $\bm{s}$. Formally,
\begin{align}
\nonumber
  e^*(t,\bm{s}) &= \int_{(0,t)\times W\times \mathcal{M}} \id_{(0,\varepsilon]}(t-\tilde{t}) \, \id_{[0,\delta]}(\norm{\bm{s}-\tilde{\bm{s}}}) \: e^{\eta_j} \, g(t-\tilde{t}) \, f(\bm{s}-\tilde{\bm{s}}) \;N(\mathrm{d}\tilde{t} \times \mathrm{d}\tilde{\bm{s}} \times \mathrm{d}\tilde{\bm{m}})\:,\\
\label{eq:twinstim-e}
&=\sum_{j\in I^*(t,\bm{s})} e^{\eta_j} \, g(t-t_j) \, f(\bm{s}-\bm{s}_j)\:,
\end{align}
where $\mathcal{M}$ is the mark space, $N$ is the time-space-mark
point process counting the infections and $I^*(t,\bm{s}) := \big\{ j
\in \{1,\ldots,N_g(t-)\} : \id_{(0,\varepsilon]}(t-t_j) = 1
\;\wedge\; \id_{[0,\delta]}(\norm{\bm{s}-\bm{s}_j}) = 1 \big\}$
is the history-dependent set of infectives at time $t$ and location $\bm{s}$,
where $N_g(t-) = N((0,t) \times W\times \mathcal{M})$.
In the above, the hyperparameters $\varepsilon, \delta > 0$ are
introduced as known \textit{maximum} temporal and spatial interaction
ranges. A past event only influences the process at time $t$ and
location $\bm{s}$, if both indicator functions are true, i.e.\ if it
occurred at most $\varepsilon$ time units ago at a location within
distance $\delta$.

\subsection{Characteristics of the Model}

Altogether, the proposed CIF model for a self-exciting spatio-temporal
point process with components \eqref{eq:twinstim-h-grid} and
\eqref{eq:twinstim-e} is
\begin{equation*} %\label{eq:twinstim}
	\lambda^*(t,\bm{s})
	=
	\offset \exp\left( \bm{\beta}'\bm{z}_{\tau(t),\xi(\bm{s})} \right)
	+
	\sum_{j\in I^*(t,\bm{s})} e^{\eta_j} \, g(t-t_j) \, f(\bm{s}-\bm{s}_j) \:,
\end{equation*}
% and we assign to the CIF model in \eqref{eq:twinstim} the name
which we shall call \texttt{twinstim} to indicate a
\emph{two}-component \emph{s}patio-\emph{t}emporal (conditional)
\emph{i}ntensity \emph{m}odel. For the proposed model an interesting
quantity is the individual-specific mean number $\mu_j$ of infections
caused by individual $j$ inside its spatio-temporal range of
interaction:
\begin{eqnarray}
    \mu_j &=& \int_0^\infty \int_{\R^2} e_j(t,\bm{s}) \,\id_{(0,\varepsilon]}(t-t_j)\, \id_{[0,\delta]}(\norm{\bm{s}-\bm{s}_j}) \dif t \dif \bm{s} \nonumber\\
    &=& e^{\eta_j} \cdot \int_0^\varepsilon g(t) \dif t \cdot \int_{b(\bm{0},\delta)} f(\bm{s}) \dif\bm{s} \:.\label{eq:twinstim:mu}
\end{eqnarray}
Here, $b(\bm{0},\delta)$ denotes the disc centred at (0,0)' with radius
$\delta$. The integration domain $\R_+\times\R^2$ above stems from the
theoretical point of view that the point process occurs in unlimited
time and space. In practice this is not observable, but
individuals near the border would be attributed a truncated value of
$\mu_j$ if integrating over $W$  -- or, similarly, $[0,T]$ -- only.
%@short2@
%resulting from the integration over $W$ instead of
%$\R^2$. Similarly, an individual which has been infected just before
%the end of the observation period at time $T$ would have $\mu_j
%\approx 0$ if only integrating over $[0,T]$.
% These edge effects are thus overcome
Such edge effects are overcome by \eqref{eq:twinstim:mu}, which also
simplifies interpretation by providing a quantity similar to the
basic reproduction number $R_0$ known from classical epidemic
modelling.
%@short@ ~\citep{anderson_may91}.
Specifically, the number $\mu_j$ offers an intuitive way of
interpreting the parameters $\bm{\gamma}$ in the linear predictor
$\eta_j$, because they
%@short2@
%An ``intercept'' term $\gamma_0$ multiplied by the two
%integral values would represent the mean number of infections caused
%by an infective individual whose marks $\bm{m}_j$ all equal zero.
%The effects of the marks
can be handled as usual in Poisson
regression models: a unit positive change in a specific continuous mark $m_{jl}$
multiplies the mean number of infections by the corresponding parameter
$e^{\gamma_l}$.

\subsection{Extension: Type-Specific \texttt{twinstim}} \label{sec:twinstim-marked}

%@short2%
% The IMD data actually represent a spatio-temporal point pattern marked
%by the finetype of infection.
Although the model of the previous subsection allows for a
finetype-specific infectivity through the vector of unpredictable marks
$\bm{m}_j$, it is not applicable for a joint modelling of both
finetypes. This is because finetypes do not change during transmission.
%@short2@ The main issue is that only the same finetype is
% transmitted but both finetypes should, however, have the same relation
% to the population at risk and, as we assume for simplicity, to the
% numbers of influenza cases and the time trend. The only
% finetype-specific element in the endemic component would be the
% intercept, corresponding to the global background rate.
% %@short@  Taking everything into account,
Therefore, the point process model will be extended to a
marked version suitable for the specific application of IMD
and point patterns with different event types in general.

Denote by $\mathcal{K} = \{1,\ldots,K\} \subset \N$ the set of
possible event types. Define an indicator matrix
%@short@\begin{equation*}
%	\bm{Q} = (q_{k,l})_{k,l\in\mathcal{K}} \qquad q_{k,l} \in \{0;1\}
%@short@\end{equation*}
$\bm{Q} = (q_{k,l})_{k,l\in\mathcal{K}}$, where $q_{k,l} \in \{0;1\}$, which
determines the possible ways of transmission. If $q_{k,l}$ equals 1,
an infective type $k$ event can cause an event of type $l$. For
instance, the IMD data would require $\bm{Q} = \bm{I}_2$, because the
transmission is finetype-specific. A marked spatio-temporal point
process on $(0,T]\times W\times \mathcal{K}$ is then defined
by the following model for the CIF:
\begin{eqnarray}
	\lambda^*(t,\bm{s},\kappa) &=& h(t,\bm{s},\kappa) + e^*(t,\bm{s},\kappa) \label{eq:twinstim:marked}\\
	h(t,\bm{s},\kappa) &=& \offset \,\exp\big( \beta_0(\kappa) + \bm{\beta}'\bm{z}_{\tau(t),\xi(\bm{s})} \big) \nonumber\\
	e^*(t,\bm{s},\kappa) &=& \sum_{j\in I^*(t,\bm{s},\kappa)} e_j(t,\bm{s}) \nonumber\\
	e_j(t,\bm{s}) &=& \exp(\eta_j) \cdot g(t-t_j\vert\kappa_j)\cdot f(\bm{s}-\bm{s}_j\vert\kappa_j) \nonumber\\
	I^*(t,\bm{s},\kappa) &=& \big\{ j \in \{1,\ldots,N_g(t-)\} :\; \id_{(0,\varepsilon]}(t-t_j) = 1 \;\wedge\; \id_{[0,\delta]}(\norm{\bm{s}-\bm{s}_j}) = 1 \;\wedge\; q_{\kappa_j,\kappa} = 1 \big\} \:.\nonumber
\end{eqnarray}
Here, the transmission indicators from the matrix $\bm{Q}$ have been
integrated into $I^*(t,\bm{s},\kappa)$.
%, where $N_g(t-)$ is the overall
%number of infections just before time $t$, and $\eta_j =
%\bm{\gamma}'\bm{m}_j$ is a linear predictor based on the event marks
%from the history of the process.
Note that the event type $\kappa_j$
is now part of the vector $\bm{m}_j$, which enables type-specific
epidemic intercepts as well as type interactions with individual
covariates in the linear predictor $\eta_j$.  The new endemic intercept $\beta_0(\kappa)$
either represents a type-specific endemic intercept, i.e.\
%\begin{equation*}
  $\beta_0(\kappa) = \sum_{k=1}^K \beta_{0,k} \,\id_{\{k=\kappa\}}(\kappa) = \beta_{0,\kappa}$,
%\end{equation*}
or contains only a single global intercept $\beta_0(\kappa)
= \beta_0$, corresponding to the hypothesis \mbox{$\beta_0 =
  \beta_{0,1} = \cdots = \beta_{0,K}$}.
%@biometrics shortening
%In any case, the parameter
%vector $\bm{\beta}$ of $h(t,\bm{s},\kappa)$ must not contain an
%intercept.
For the remainder of the endemic predictor, the model
assumes independence of $\kappa$, which means that the effect of
endemic covariates is homogeneous over the event types.  However, the
history-dependent set $I^*(t,\bm{s},\kappa)$ of infective individuals
now accounts for the transmission regime $\bm{Q}$ between the event
types, and the interaction functions are allowed to depend on the type
of the infective event as well.
%@short2%
% For instance by using type-specific $\sigma_\kappa^2$
% %@short@ and $\alpha_\kappa$ in \eqref{eq:f-gaussian} and
% %\eqref{eq:g-expdecay}, respectively.
% and $\alpha_\kappa$ in the spatial and temporal interaction functions,
% respectively.

\section{Statistical Inference}
\label{sec:inference}
This section deals with likelihood inference for the parameters of the
CIF in \eqref{eq:twinstim:marked} based on the observed marked
spatio-temporal point pattern $\bm{x} = \{(t_i, \bm{s}_i,\bm{m}_i):
i=1,\ldots,n\}$, where the event type $\kappa_i$ is part of the vector
of marks $\bm{m}_i$, and $n$ is the number of events, i.e.\ a
realisation of $N_g(T)$.  The parameter vector in question is
%@short@ \begin{equation} \label{eq:theta}
%	\bm{\theta} = (\bm{\beta_0}', \bm{\beta}', \bm{\gamma}', \bm{\sigma}', \bm{\alpha}')' \:,
%@short@ \end{equation}
$\bm{\theta} = (\bm{\beta_0}', \bm{\beta}', \bm{\gamma}',
\bm{\sigma}', \bm{\alpha}')'$, where
%@revision
%$\bm{\beta_0} =
%(\beta_{0,1},\ldots,\beta_{0,K})'$ (type-specific) or $\bm{\beta_0} =
%\beta_0$ (type-invariant), and
$\bm{\sigma}$ and $\bm{\alpha}$ are the parameter vectors of the
spatial and temporal interaction functions $f_{\bm{\sigma}}$ and
$g_{\bm{\alpha}}$, respectively.

%In a frequentist framework, parameter estimates can be obtained by
%maximisation of the log-likelihood %@short2@ or the partial log-likelihood
%with respect to $\bm{\theta}$.
%@short2@
% Trading the partial likelihood off against the full likelihood, the
% results of the simulation study in \citet{Diggle2009} support
% satisfactory relative efficiency of the partial likelihood for their
% model, which can be regarded as a special instance of
% \texttt{twinstim}. However, we do not see great benefit in using the
% partial likelihood approach because the need for spatial integration
% -- which here is the computational bottleneck of statistical inference
% -- remains.  Furthermore, the parameters $e^{\beta_0}$ and
% $e^{\gamma_0}$ would not both be identifiable, but only their
% ratio. As a consequence, we will concentrate on full maximum
% likelihood estimation.
% %@short@-- although our implementation also provides
% %the partial likelihood alternative.
% In the next subsections, the log-likelihood and score functions
% related to the type-specific \texttt{twinstim} are
% derived. Furthermore, estimation of the expected Fisher information
% matrix and asymptotic properties of the maximum likelihood
% estimators are discussed.

%\subsection{Log-Likelihood Function} \label{sec:twinstim:loglik}

In our framework, no attempt is made to model unpredictable marks
like gender and age but they are taken as given predictor variables in
models of the CIF. In this case, the log-likelihood of
the underlying point process $N$ on $[0;T]\times W\times \mathcal{M}$ may be
conveniently written as~\citep{DV2003}
\begin{equation*} %\label{eq:loglik-ST-marked-decoupled}
%  \left[
    \sum_{i=1}^n \log\lambda_{\bm{\theta}}^*(t_i, \bm{s}_i, \kappa_i)
    - \int_0^T \int_W \sum_{\kappa\in \mathcal{K}}
    \lambda_{\bm{\theta}}^*(t,\bm{s},\kappa) \dif t \dif\bm{s} \:.
%  \right]
\end{equation*}

The components of the above sum can be directly calculated for a specific
value of the parameter vector $\bm{\theta}$ after having determined
the set $I^*(t_i,\bm{s}_i,\kappa_i)$ of potential sources of infection
for the $i$th event. Furthermore,
%@biometrics shortening
%the integrated conditional intensity function in the log-likelihood is
%\begin{equation*}
%\int_0^T \int_W \sum_{\kappa \in \mathcal{K}} \lambda_{\bm{\theta}}^*(t,\bm{s},\kappa) \dif t \dif\bm{s}
%= \int_0^T \int_W \sum_{\kappa \in \mathcal{K}} h_{\bm{\theta}}(t,\bm{s},\kappa) \dif t \dif\bm{s} + \int_0^T \int_W \sum_{\kappa \in \mathcal{K}} e_{\bm{\theta}}^*(t,\bm{s},\kappa) \dif t \dif\bm{s}
%\end{equation*}
the integrations of the endemic and epidemic components of the CIF
can be performed separately due to their additive superposition.
Recalling that the endemic component is a
piecewise constant function on the spatio-temporal grid
$\{C_1,\ldots,C_D\} \times \{A_1,\ldots,A_M\}$, its integral is
in fact a sum over this grid of smallest observed units in space-time:
\begin{equation} \label{eq:h-integral}
\int_0^T \int_W \sum_{\kappa \in \mathcal{K}} h_{\bm{\theta}}(t,\bm{s},\kappa) \dif t \dif\bm{s} =
\Big( \sum_{\kappa \in \mathcal{K}} \exp\left(\beta_0(\kappa)\right) \Big) \cdot
\sum_{\tau=1}^D \sum_{\xi=1}^M \abs{C_\tau}\abs{A_\xi} \rho_{\tau,\xi} \exp\left( \bm{\beta}'\bm{z}_{\tau,\xi} \right) \:.
\end{equation}

The integrated epidemic component can be simplified by moving the
indicators of the function $I^*(t,\bm{s},\kappa)$ back into the sum:
\begin{align}
&\int_0^T \int_W \sum_{\kappa \in \mathcal{K}} e_{\bm{\theta}}^*(t,\bm{s},\kappa) \dif t \dif\bm{s} \nonumber\\
	&=\int_0^T \int_W \sum_{\kappa \in \mathcal{K}} \sum_{j=1}^n
	\id_{(0,\varepsilon]}(t-t_j) \,\id_{[0,\delta]}(\norm{\bm{s}-\bm{s}_j}) \, q_{\kappa_j,\kappa}\: e^{\eta_j} \, g_{\bm{\alpha}}(t-t_j\vert\kappa_j)\, f_{\bm{\sigma}}(\bm{s}-\bm{s}_j\vert\kappa_j) \,\dif t \dif\bm{s} \nonumber\\
%	&= \sum_{j=1}^n q_{\kappa_j,\text{\tiny\textbullet}} \:e^{\eta_j} \bigg[\int_0^T \id_{(0,\varepsilon]}(t-t_j) g_{\bm{\alpha}}(t-t_j\vert\kappa_j) \dif t\bigg] \bigg[ \int_W \id_{[0,\delta]}(\norm{\bm{s}-\bm{s}_j}) f_{\bm{\sigma}}(\bm{s}-\bm{s}_j\vert\kappa_j) \dif\bm{s} \bigg] \nonumber\\
	&= \sum_{j=1}^n q_{\kappa_j,\text{\tiny\textbullet}} \:e^{\eta_j} \Big( \int_0^{\min\{T-t_j;\varepsilon\}} g_{\bm{\alpha}}(t\vert\kappa_j) \dif t \Big) \Big( \int_{R_j} f_{\bm{\sigma}}(\bm{s}\vert\kappa_j) \dif\bm{s} \Big) \:. \label{eq:e-integral}
\end{align}
Here, $q_{\kappa_j,\text{\tiny\textbullet}} := \sum_{\kappa \in
    \mathcal{K}} q_{\kappa_j,\kappa}$ is the number of
different event types that can be triggered by an event of type
$\kappa_j$, and
%\begin{equation} \label{eq:influenceRegion}
	$R_j := \big\{ W\cap b(\bm{s}_j;\delta) \big\} - \bm{s}_j$
%\end{equation}
is the spatial interaction region of the $j$th event centred at its location.
%@short biometrics
%In the case of unlimited spatial
%transmission ($\delta=\infty$), $R_j = W-\bm{s}_j$ equals the
%translation of the whole observation region by $\bm{s}_j$ such that
%$\bm{s}_j$ becomes the origin.

The evaluation of the two-dimensional integral over the domains $R_j$
is the most sophisticated task of the log-likelihood
evaluation. \citet{Meyer2009} compared accuracy and speed of different
cubature rules for performing the numerical integration. Here, the
two-dimensional midpoint rule~\citep[see e.g.][]{stroud71} proved to
be best suited for the task.
%For the special case of the
%%Gaussian kernel \eqref{eq:f-gaussian}, robust accuracy for any value
%Gaussian kernel, robust accuracy for any value of $\sigma$ can be
%achieved by an adaptive choice of the bandwidth $h=\phi\,\sigma$.
%@short@ using an appropriate accuracy $\phi$, see \citet[3.2.7]{Meyer2009} for details.
In contrast, the evaluation of the definite integral over the temporal
interaction function is analytically accessible for typical choices of
$g_{\bm{\alpha}}$.
%@short practice
% Provided $G_{\bm{\alpha}}(t\vert\kappa)$ denotes an antiderivative of
% $g_{\bm{\alpha}}(t\vert\kappa)$, the first integral thus equals
% \begin{equation*}
% 	\int_0^{\min\{T-t_j;\varepsilon\}} g_{\bm{\alpha}}(t\vert\kappa_j) \dif t = G_{\bm{\alpha}}\big(\min\{T-t_j;\varepsilon\}\big\vert\kappa_j\big) - G_{\bm{\alpha}}\big(0\big\vert\kappa_j\big) \:.
% \end{equation*}
% For instance, the type-specific exponential decay function
% %@short@ $g_{\bm{\alpha}}$ from equation \eqref{eq:g-expdecay} has
% $g_{\bm{\alpha}}$ has antiderivative
% %\begin{equation} \label{eq:G-expdecay}
% %	G_{\bm{\alpha}}(t\vert\kappa) = -\frac{e^{-\alpha_\kappa\,t}}{\alpha_\kappa} \qquad (\alpha_\kappa > 0)\:.
% %\end{equation}
% $G_{\bm{\alpha}}(t\vert\kappa) =
% -e^{-\alpha_\kappa\,t}/\alpha_\kappa$, $\alpha_\kappa >
% 0$.
%@short2@
%The case $\alpha_\kappa = 0$ would correspond to a time-invariant
%infectivity, i.e.\ $g_{\bm{\alpha}}(t\vert\kappa) = 1$ with
%antiderivative $G_{\bm{\alpha}}(t\vert\kappa) = t$.

Altogether, an analytical maximisation of the above log-likelihood is
not feasible, and a numerical optimisation routine such as
BFGS~\citep[see e.g.][Section 8.1]{nocedal_wright99} is required.
Here, it is advantageous to know the score function $s(\bm{\theta})$,
which is derived in Web Appendix A. Uncertainty of the parameter
estimates is deduced from the expected Fisher information
$\mathcal{I}(\bm{\theta})$ as estimated by the ``optional variation
process'' adapted to the marked spatio-temporal setting -- see Web
Appendix B
%@biometrics shortening and \citet[equation (4.7)]{Rathbun1996}
for details. Significance of specific model parameters can be
investigated by Wald or likelihood ratio tests and model selection is
performed based on Akaike's information criterion (AIC).

\section{Simulation Algorithm}
\label{sec:simulation}
In general, the usability of a model class is greatly improved by the ability to simulate from a specific model. For instance, it enables model checking and parametric bootstrap.
For evolutionary point processes specified by their CIF, \emph{Ogata's modified thinning algorithm} \citep[Algorithm 7.5.V.]{DV2003} provides a convenient and exact way to simulate realisations of the process. The algorithm requires piecewise upper bounds for the intensity $\lambda^*_g(t)$ of the ground process $N_g(t) := N((0,t]\times W\times\mathcal{K})$.
%@biometrics shortening
%, which counts the total number of points occuring up to time $t$ anywhere in $W$ and of any type $\kappa \in \mathcal{K}$.
This intensity is determined as
\begin{eqnarray*}
	\lambda^*_g(t) &=& \int_W \sum_{\kappa \in \mathcal{K}} \lambda^*(t,\bm{s},\kappa) \dif\bm{s} = \left( \sum_{\kappa \in \mathcal{K}} e^{\beta_0(\kappa)} \right) \left( \sum_{\xi=1}^M \abs{A_\xi} \, \rho_{\tau(t),\xi}\, e^{\bm{\beta}'\bm{z}_{\tau(t),\xi}} \right) \\
	&&{}+\; \sum_{j=1}^{N_g(t-)} \left( \sum_{\kappa\in\mathcal{K}} q_{\kappa_j,\kappa} \right) \, e^{\eta_j} \,\id_{(0,\varepsilon]}(t-t_j) \, g(t-t_j\vert\kappa_j) \, \int_{R_j} f(\bm{s}\vert\kappa_j) \dif\bm{s} \:.
\end{eqnarray*}
This function is bounded above by the CIF $\overline{\lambda_g^*}(t)$,
which is defined by replacing $g(t\vert\kappa)$ by the \emph{constant}
temporal interaction function $\overline{g}(t\vert\kappa) =
\max\limits_{u>0} g(u\vert\kappa)$. This CIF is piecewise constant in
time as it only jumps at time points where any of the endemic
covariates in $\bm{z}_{\tau(t),\xi}$ in any tile $\xi$ changes its
value, or when the set of currently infectious individuals changes,
i.e.\ whenever a new event occurs or a previous event stops triggering.

Given a parameter vector $\bm{\theta}$, the ranges of interaction
$\varepsilon$ and $\delta$, as well as a sampling scheme for the marks
$\bm{m}_j$, the time point of the next infection starting from the
current time $t=t_0$ can be generated as follows: Draw an
exponentially distributed random variate $\Delta$ with rate
$\overline{\lambda_g^*}(t_0)$. The simulated value of $\Delta$ is a
proposal for the waiting time to the next event, i.e.\ the next time
point of infection might be $\tilde{t} = t_0 + \Delta$. However, this
proposal is not valid if the rate $\overline{\lambda_g^*}(t)$ had
changed between $t_0$ and $\tilde{t}$. In this case, time is set to
the first changepoint after $t_0$ and a new $\Delta$ is
simulated. Eventually, a proposed time point $\tilde{t}$ is valid. It
is then accepted with probability $\lambda_g^*(\tilde{t}) /
\overline{\lambda_g^*}(\tilde{t})$. If it is rejected, time is set to
$t=\tilde{t}$ and a new waiting time $\Delta$ is simulated as
above. If it is accepted, location $\bm{\tilde{s}}$ and type
$\tilde{\kappa}$ of the event have to be simulated. At first, the
source of infection is sampled with probabilities proportional to the
respective components of $\lambda_g^*(\tilde{t})$:
\begin{eqnarray}
\P(\text{endemic source}) \cdot \lambda_g^*(\tilde{t}) &=& \left( \sum_{\kappa \in \mathcal{K}} e^{\beta_0(\kappa)} \right) \left( \sum_{\xi=1}^M \abs{A_\xi} \,\rho_{\tau(\tilde{t}),\xi} \, e^{\bm{\beta}'\bm{z}_{\tau(\tilde{t}),\xi}} \right) \label{eq:sourceprobs}\\
\P(\text{source} = \text{event } j) \cdot \lambda_g^*(\tilde{t}) &=& \left( \sum_{\kappa\in\mathcal{K}} q_{\kappa_j,\kappa} \right) \, e^{\eta_j} \,\id_{(0,\varepsilon]}(\tilde{t}-t_j) \, g(\tilde{t}-t_j\vert\kappa_j) \, \int_{R_j} f(\bm{s}\vert\kappa_j) \dif\bm{s}\:, \nonumber
\end{eqnarray}
for $j \in \{1,\ldots,N_g(\tilde{t}-)\}$. On the one hand, if the new event has an endemic source, then
% \begin{eqnarray*}
% 	\P(\tilde{\kappa} = k) &\propto& \exp(\beta_0(k)) \qquad(k\in\mathcal{K}) \\
% 	\P(\bm{\tilde{s}} \in A_\xi) &\propto& \abs{A_\xi} \,\rho_{\tau(\tilde{t},\xi}\, e^{\bm{\beta}'\bm{z}_{\tau(\tilde{t}),\xi}} \qquad(\xi=1,\ldots,M)
% \end{eqnarray*}
$\P(\tilde{\kappa} = k) \propto \exp(\beta_0(k))$, $k\in\mathcal{K}$,
and $\P(\bm{\tilde{s}} \in A_\xi) \propto \abs{A_\xi} \,\rho_{\tau(\tilde{t}),\xi}\,e^{\bm{\beta}'\bm{z}_{\tau(\tilde{t}),\xi}}$, $\xi=1,\ldots,M$. In the
sampled tile $A_{\tilde{\xi}}$, the location $\bm{\tilde{s}}$ is
uniformly distributed.
On the other hand, if the new event was triggered by the previous
event $j$, then $\tilde{\kappa} \sim U(\{k: q_{\kappa_j,k} = 1\})$,
%@short2@
%(i.e.\ one of the types which can be triggered by the type $\kappa_j$
%without any weighting),
and $\bm{\tilde{s}} = \bm{s}_j + \bm{v}$, where $\bm{v}$
is drawn from the density $f(\bm{s}\vert\kappa_j) / \int_{R_j}
f(\bm{s}\vert\kappa_j) \dif s$ on $R_j$, e.g.\ using rejection
sampling.

A scheme of the described algorithm can be found as Web Appendix C.

\section{Application to the IMD Data}
\label{sec:application}

Although visual comparisons between the finetypes and heuristic
comparisons of the estimates of separate finetype-specific models are
possible, this does not allow to assess potential differences
statistically. We thus conduct a joint analysis of the two finetypes
by the marked \texttt{twinstim} of Section~\ref{sec:twinstim-marked}.
%@short@
%in order to test whether the weight of the epidemic component and hence
%the basic reproduction number is significantly higher for the serogroup
%B finetype.
%%%%Similarly, testing for a different global endemic
%%%%baseline risk $\exp(\beta_0)$ is investigated.
We perform model selection for the joint point pattern of
630 cases of IMD with complete age and gender
information by using AIC to compare
all models with the CIF composed by subsets of the following terms:
\begin{itemize}
\item Endemic component: common or finetype-specific intercept, linear
  time trend, time-of-year effects (one or two harmonics), and
  linear effect of weekly number of influenza cases registered in the
  district of a point (no time lag, lags 0 and 1, lags 0--2, or lags 0--3)
  taken from the SurvStat database~\citep{RKI-survstat}.
\item Epidemic component: gender, age (categorized as 0-2, 3-18 and $\geq\! 19$ years), finetype, and age-finetype interaction.
\end{itemize}
As an offset in the endemic component, we use
the district-specific population density $\rho_{\xi(\bm{s})}$
(inhabitants per km$^2$).
A fixed hyperparameter of $\varepsilon=30$ days is assumed --
  this maximal temporal interaction range is consistent with the range
  used in, e.g., \citet{zangwill_etal97}. Because the number of supposedly direct
  transmissions in the IMD dataset is humble, we will furthermore
  assume a constant temporal interaction function $g$ (i.e.\ constant
  spread within the $\varepsilon$ days) in order to not
  overparametrize the epidemic component. The spatial
  hyperparameter is fixed at $\delta=200$ km -- this parameter needs
  only to be large enough not to influence the estimation of the
  actual spatial interaction function $f$.

To restrict the model search, and hence computing time, we first
performed the search for all 600 models
($2\cdot 2\cdot 3\cdot 5$ configurations of the endemic component and
$2\cdot 5$ configurations of the epidemic component)
with constant spatial
interaction function $f$. Hereafter, the top 10 models of this search
were investigated further with two Gaussian spatial interaction
functions: one with joint variance parameter and one with
finetype-specific variance parameter.

The CIF of the resulting AIC-best model obtained by this search was $\lambda_{\bm{\theta}}^*(t,\bm{s},\kappa) =$
\begin{eqnarray*} %\label{eq:joint-model0}
&& \rho_{\xi(\bm{s})}
  \cdot \exp\Big(\beta_{0} + \beta_{\text{trend}}
  \tfrac{\lfloor t\rfloor}{365} + \beta_{\sin} \sin\big(\lfloor
  t\rfloor\,\tfrac{2\pi}{365}\big) + \beta_{\cos} \cos\big(\lfloor
  t\rfloor\, \tfrac{2\pi}{365}\big) \Big) \\
&+& \!\!\sum_{j\in
    I^*(t,\bm{s},\kappa)} q_{\kappa_j,\kappa} \;
  \exp\left( \gamma_{0} + \gamma_{\text{3-18}} \id_{[3,18]}(\text{age}_j) + \gamma_{\geq 19} \id_{[19,\infty)}(\text{age}_j) + \gamma_{\text{C}} \id_{\{C\}}(\kappa_j) \right) \; f_{\sigma}(\bm{s}-\bm{s}_j).
\end{eqnarray*}
Here, $(t,\bm{s},\kappa)$ denotes days since 31 December 2001,
coordinate in ETRS89 (kilometre scale) and finetype. With $\lfloor
t\rfloor$ we denote monday of week $\tau(t)$, i.e.\ the lower bound of
time intervals $C_1,\ldots,C_D$.
In the linear predictor of the epidemic component, age group
0-2 and type B serve as reference categories. The corresponding
parameter estimates of the best model, now fitted to the
635 cases with available age, are found in
Table~\ref{tab:aicmodel}.

\begin{table}
\caption{Parameter estimates for the endemic (top) and epidemic (bottom)
component of the model with the lowest AIC (AIC=18968).
%with \texttt{h.<NAME>} denoting endemic and \texttt{e.<NAME>} denoting epidemic components,
%  e.g.\ \texttt{e.siaf.B}=$\log \sigma_B$.
%  e.g.\ \texttt{e.siaf}=$\log \sigma$.
  The $p$-values correspond to Wald tests.}
\label{tab:aicmodel}
\begin{center} \small
\begin{tabular}{rrrrr}
\hline
 & Estimate & Std. Error & $z$ value & $\P(|Z|>|z|)$ \\
\hline
\hline
% latex table generated in R 2.13.1 by xtable 1.5-6 package
% Mon Oct  3 16:53:56 2011
 $\beta_0$ & $-20.3652$ & $0.0872$ & $-233.53$ & $<2\cdot{}10^{-16}$ \\ 
  $\beta_{\text{trend}}$ & $-0.0493$ & $0.0223$ & $-2.21$ & $0.027$ \\ 
  $\beta_{\sin}$ & $0.2618$ & $0.0649$ & $4.03$ & $5.5\cdot{}10^{-05}$ \\ 
  $\beta_{\cos}$ & $0.2668$ & $0.0644$ & $4.14$ & $3.4\cdot{}10^{-05}$ \\ 
  \hline
% latex table generated in R 2.13.1 by xtable 1.5-6 package
% Mon Oct  3 16:53:56 2011
 $\gamma_0$ & $-12.5746$ & $0.3128$ & $-40.21$ & $<2\cdot{}10^{-16}$ \\ 
  $\gamma_{\text{3-18}}$ & $0.6463$ & $0.3195$ & $2.02$ & $0.04310$ \\ 
  $\gamma_{\geq 19}$ & $-0.1868$ & $0.4321$ & $-0.43$ & $0.66558$ \\ 
  $\gamma_{\text{C}}$ & $-0.8496$ & $0.2574$ & $-3.30$ & $0.00097$ \\ 
  $\log\sigma$ & $2.8287$ & $0.0819$ &  &  \\ 
  \hline
\hline
\end{tabular}\end{center}
\end{table}

\begin{figure}%[bht]
\centering
\subfigure[]{
\includegraphics{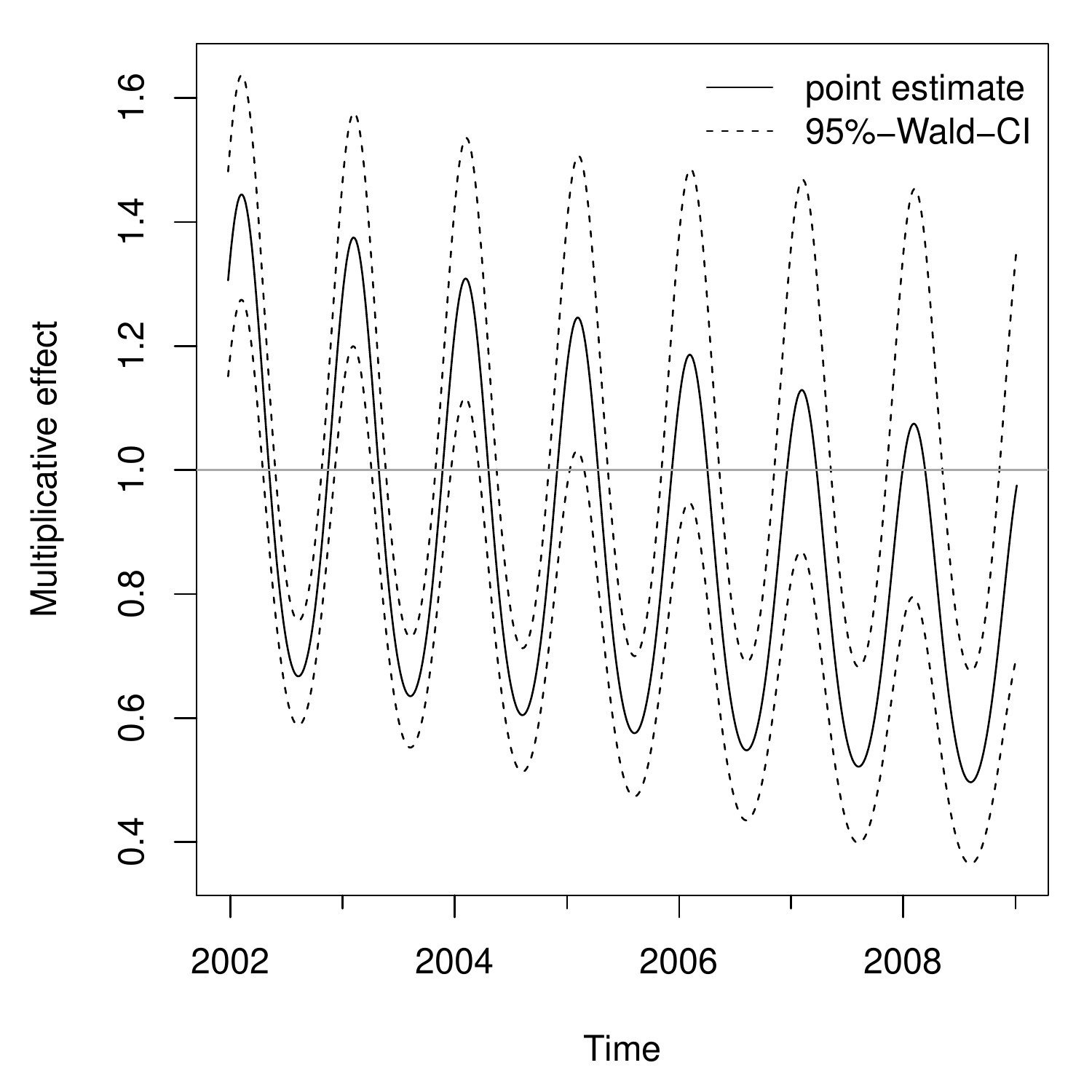}
\label{fig:IMD1}
}
\subfigure[]{
\includegraphics{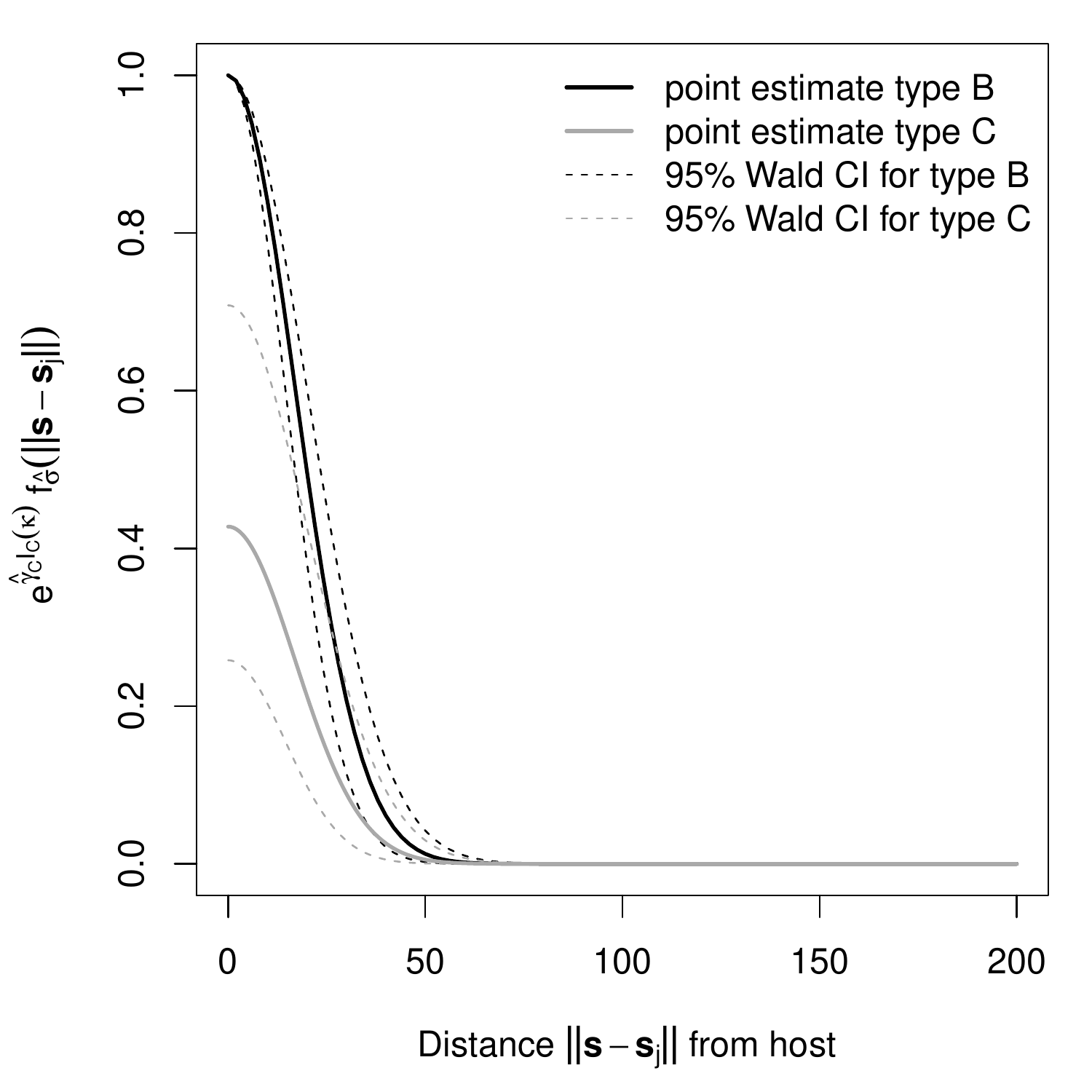}
\label{fig:IMD2}
}
\caption{(a) Trend and seasonal component of the fitted model; one
  observes the typical IMD peak in late February and minimum in
  August. Furthermore, (b) shows the spatial interaction function
  multiplied by the type modifier illustrating the higher epidemic potential of type B.}
% the effective interaction range appears to be about
%50
%  (type B) and \Sexpr{5*(u[which.min(hat.c >
%    0.01)] %/% 5)}\,km (type C),
%    respectively.}
\label{fig:results-seas-siaf}
\end{figure}

Thus, there appears to be no noteworthy difference in the endemic
behaviour of the two types: a linear downward time trend superimposed
with one harmonic best describes the endemic behaviour of the point
pattern (see Figure~\ref{fig:IMD1}). An additional effect of past
numbers of influenza cases does not improve the model. In contrast,
there is an effect of past IMD cases, i.e.\ the process is indeed
self-exciting.  Comparing the endemic-only model with the model
  enriched by an epidemic intercept only, greatly improves the fit
  ($\Delta$AIC=202.84).
%    In the epidemic component, there is no detectable
%    dependence on marks, except that the range of the spatial interaction
%    function is type-dependent with type B spreading further away (see
%    also Figure~\ref{fig:IMD2}).
    In the epidemic component, there is a detectable dependence on
    marks with type C being less aggressive than type B.
%@short@
% (i.e.\ $\hat{\gamma}_C=-0.85<0$)
    Figure~\ref{fig:IMD2} shows the resulting finetype-specific
    spatial interaction functions which for type C is
    $e^{\hat{\gamma}_C} \cdot
    100\% = 43$\% of type B.
      Finally, there is a significant age difference in the
      infectivity of cases: the highest potential is found in the 3-18
      year old, which could be interpreted as the kindergarten and
      school-aged children having a higher contact behaviour than
      e.g.\ adults.

%    \TODO{Note also, that the following three
%models have an AIC difference less than two to the AIC-best model: 1)
%two instead of one sine-cosine components, 2) type-specific epidemic
%intercept, and 3) type-specific endemic intercept.}

% Das macht keinen Spass zu untersuchen: \TODO{Considering
%  the logarithm of the numbers of influenza cases did not reveal
%  substantial changes.}

%Maybe later if reviewer requests this. Fit in m.partial, but
%do we need to address reparametrization?
%\TODO{Compare with fit using the partial likelihood approach...}

Based on the selected model, basic reproduction numbers of $\hat\mu_{\text{B}}=0.25$
(95\% CI 0.19-0.34) vs. $\hat\mu_{\text{C}}=0.11$
(95\% CI 0.07-0.17)
are obtained by calculating the type-specific expectation of
\eqref{eq:twinstim:mu} over the empirical distribution function of the
additional covariates in the epidemic predictor (here: age group).
The confidence intervals are given as the 0.025 and 0.975 quantiles of
samples obtained by re-computing $\hat\mu_{\text{B}}$ and
$\hat\mu_{\text{C}}$ for 999 additional coefficient
vectors drawn from the asymptotic multivariate normal distribution of
the parameter estimates in Table~\ref{tab:aicmodel}.  The confidence
intervals thus indicate a higher epidemic potential of the serogroup B finetype.
%new comment
Note that these numbers are lower than what one would expect from the
literature, e.g.\ \citet{trotter_etal2005} report an $R_0$ estimate of
1.36 for \textit{serogroup} C. Two explanations account for this
discrepancy: firstly, our estimation is based on transmission between
cases with invasive disease and not between asymptomatic carriers, who
are not represented in disease surveillance data.  Secondly, use of an
endemic component means that our $R_0$ estimates are destined to be
lower, because sporadic cases do not contribute to the number of
secondary cases. Still, our estimates provide realistic lower bounds
for carriage reproduction numbers.

%\item LR-test for $H_0: \gamma_{0,B}=\gamma_{0,C}$ has $p$-value $0.0011$

\setkeys{Gin}{width=0.45\textwidth}

\begin{figure}
\centering
\subfigure[$\epsilon$-scheme.]{
\includegraphics{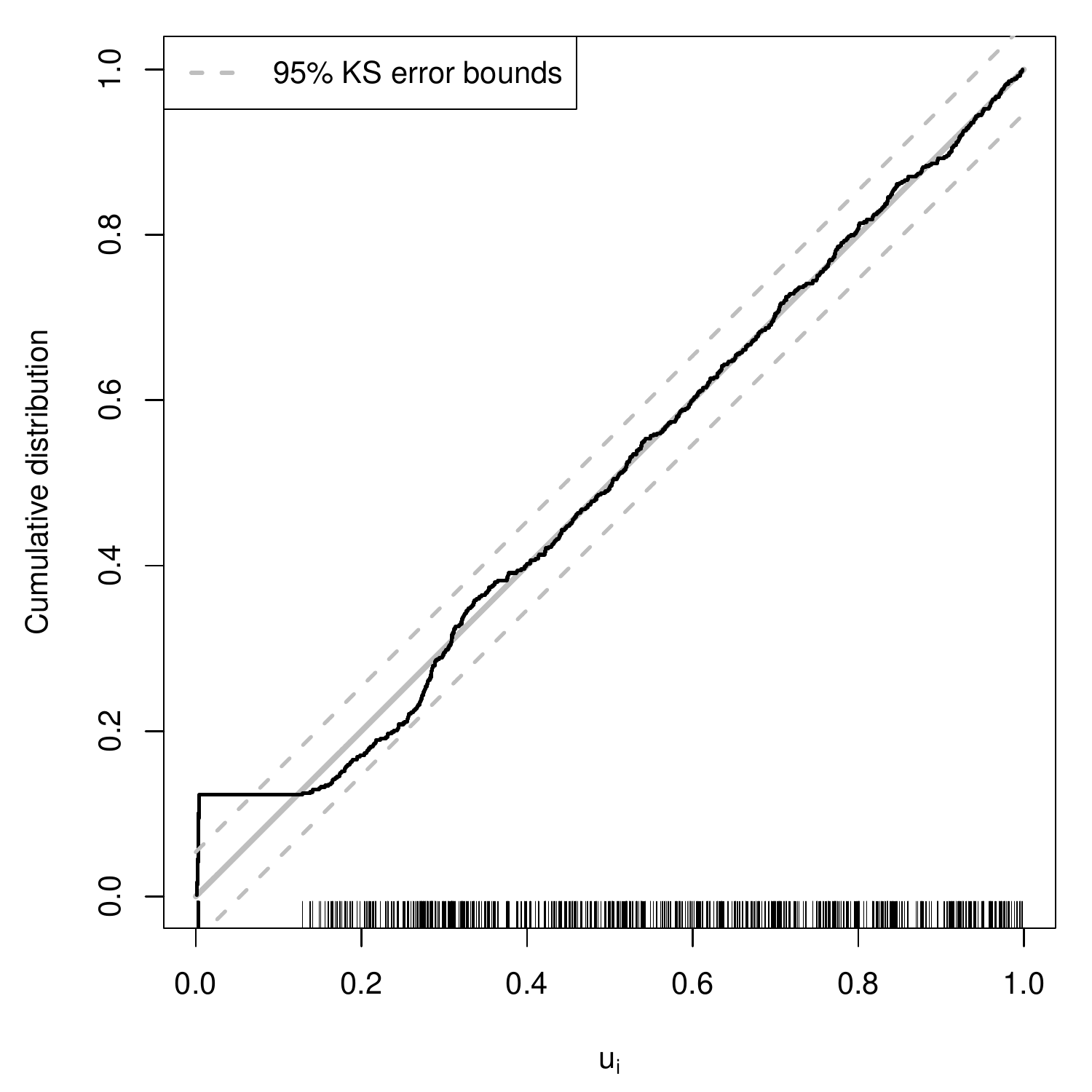}
}
\subfigure[$U(0,1)$-scheme.]{
\includegraphics{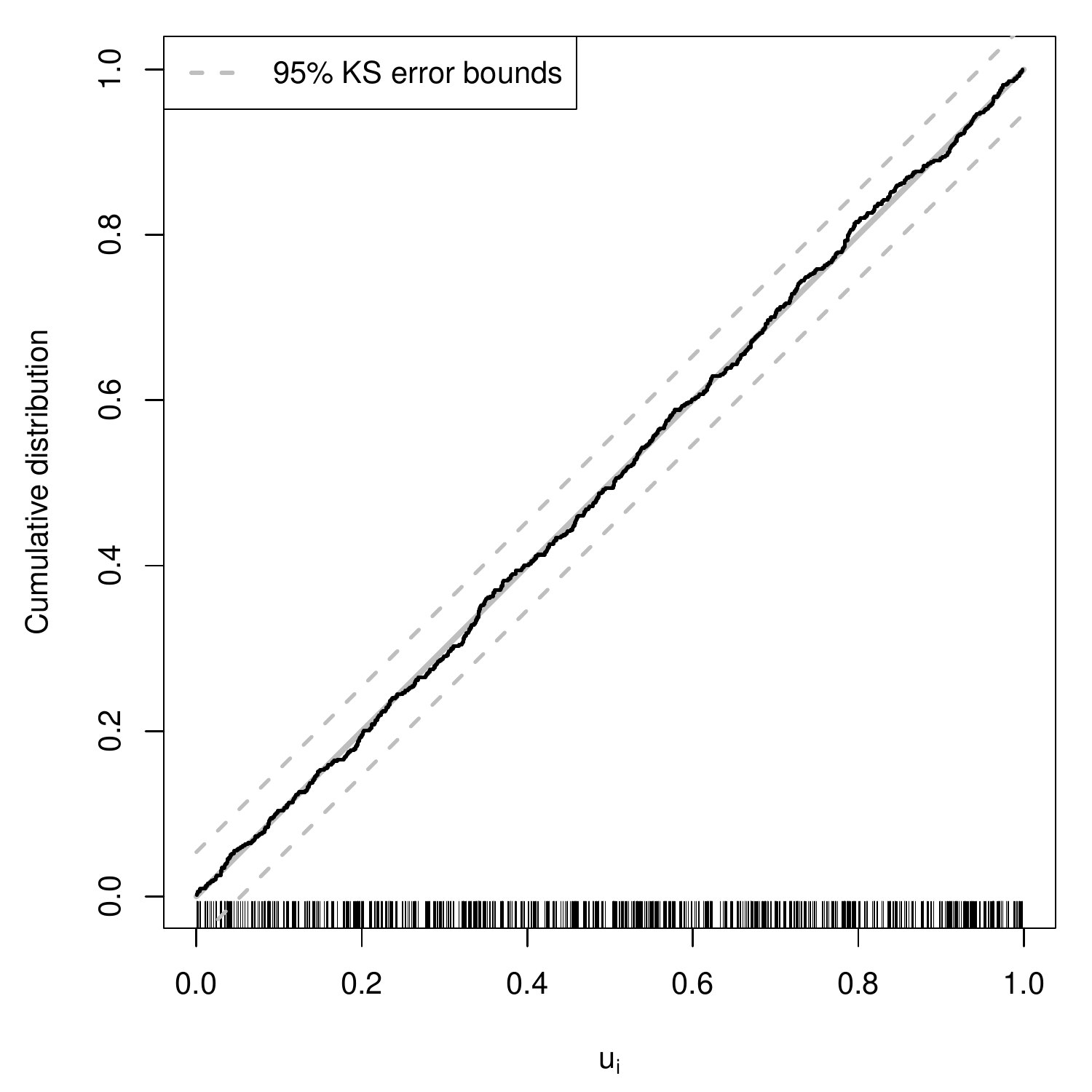}
}
  \caption{CDF of the observed $U_i$ together with 95\%
    Kolmogorov-Smirnov error bounds for data with tie breaking
    according to the (a) $\epsilon$ scheme and (b) $U(0,1)$ scheme.}
\label{fig:residuals}
\end{figure}

To inspect the goodness-of-fit of the selected
spatio-temporal point process model, we follow the suggestion by
\citet{Ogata1988} \citep[see also][]{Rathbun1996} by computing
%@short@ \begin{align*}
%   Y_i &= \hat{\Lambda}_g^*(t_i) - \hat{\Lambda}_g^*(t_{i-1}), \quad
%   i=2,\ldots,n,
% \end{align*}
$Y_i = \hat{\Lambda}_g^*(t_i) - \hat{\Lambda}_g^*(t_{i-1})$,
$i=2,\ldots,n$, where $\hat{\Lambda}_g^*(t)$ is the fitted cumulative
intensity function of the ground process.  If the estimated CIF
describes the true CIF well, then
$U_i=1-\exp(-Y_i)\stackrel{\text{iid}}{\sim}
U(0,1)$. Figure~\ref{fig:residuals}(a) contains a plot of the
cumulative density function (CDF) of the observed $U_i$ and for
comparison the CDF of the $U(0,1)$-distribution together with error
bounds computed by inverting the one sample Kolmogorov-Smirnov
test. The fit appears good, but noticable deviations for $u_i<0.15$
can be observed, which we suspect to occur due to the tie-breaking
strategy of subtracting $\epsilon=0.01$ days from ties. As
observations are on a per-day basis and thus are interval censored we
re-estimated the model for a data set where ties were broken by
subtracting a $U(0,1)$-distributed random number from each observation
time. Figure~\ref{fig:residuals}(b) shows the improved fit of this
analysis -- the relative changes in the parameter estimates are minor.
%@short%
%and in the order of
%0.98 -
%
%  1.07 except
% for the estimate
%   $\hat{\gamma}_{\geq 19}$ where the change is of order
%
%   1.43.
%      The ``residual plots'' in Figure~\ref{fig:residuals} thus
%      provided important insights with respect to the obtained fit and
%      suggest reasonable goodness-of-fit.
%      In addition, a scatterplot
%      of the observed $U_i$ against $U_{i+1}$ (not shown) yields no
%      evidence for serial correlation.

\setkeys{Gin}{width=0.45\textwidth}

\begin{figure}
\centering
\includegraphics[width=0.6\textwidth]{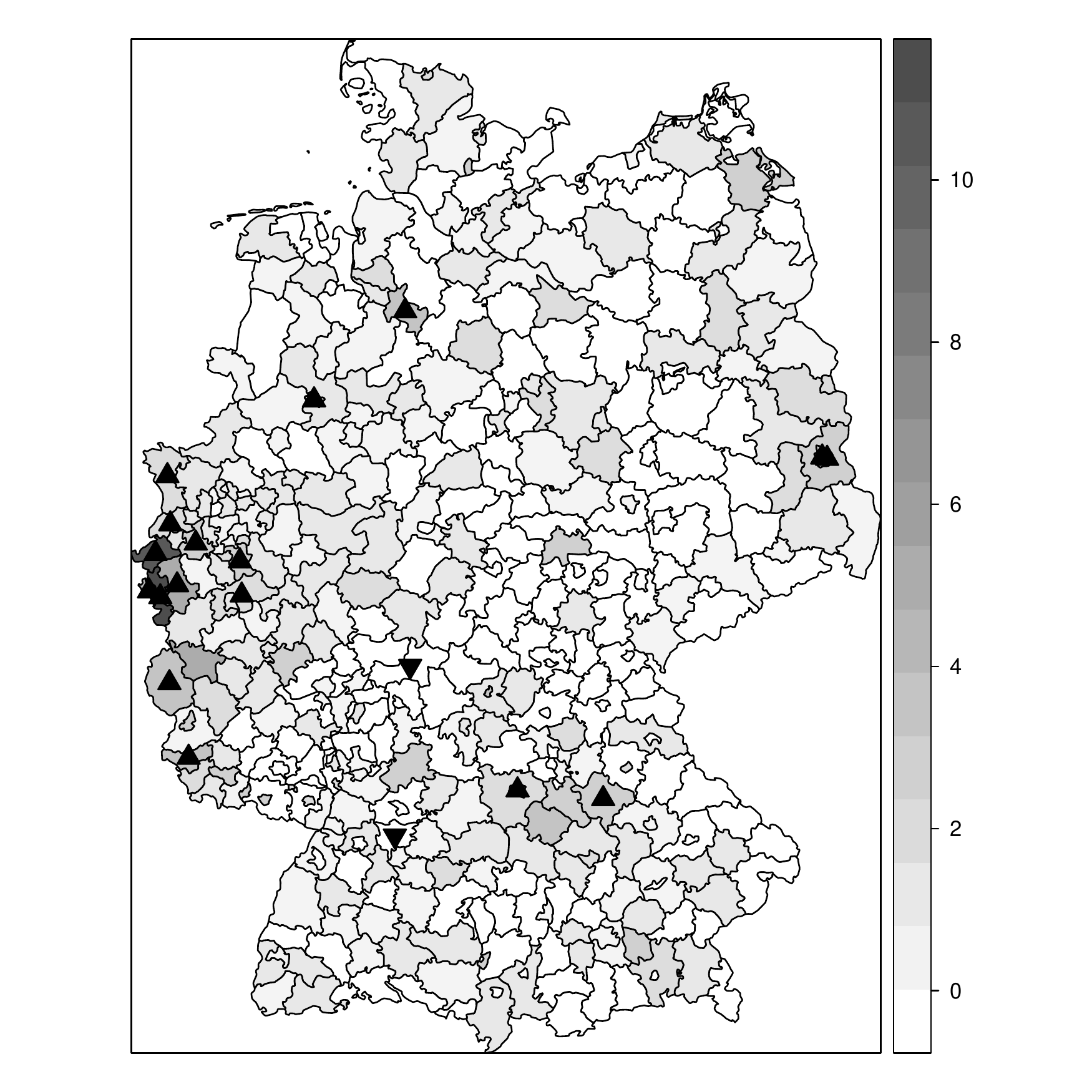}
\caption{Observed incidence (per 100,000 inhabitants) during 2002-2008 for both finetypes together. Triangles pointing up (down) indicate districts with a higher (lower) incidence than explained by 100 simulations from the model.}
\label{fig:incidence+outliers}
\end{figure}

Another way of assessing the goodness-of-fit is by simulation from the fitted CIF.
Figure~\ref{fig:incidence+outliers} shows the observed 7-year incidences (per
100,000 inhabitants) of the 413 districts for both finetypes together. In order
to identify extreme observations that are not explained by the selected model,
we simulated 100 realisations of the process
%@short@ along the lines of Section~\ref{sec:simulation},
and determined the 2.5\% and 97.5\% quantiles of the district-specific
7-year incidences. In the figure, districts with observed incidences
outside the simulated 95\%-range are marked by triangles. Many of the
17 districts with an excess are found around the city Aachen at
the border to the Netherlands. The deviation from the model could thus
be explained by edge effects hiding potential transmissions across the
border.

  Altogether, we are led to the conclusion that the proposed
  model provides a useful description of the spread of IMD. It allows
  a quantification that the serogroup B finetype has a higher epidemic
  potential than the serogroup C finetype and shows age difference in
  spread behaviour. A sensitivity analysis confirmed robustness of
  these results for increasing values of $\delta$.  Order and
  significance of the finetype difference in the epidemic component
  remained stable for $\varepsilon$ in the range of 1-5 weeks to 1-4
  months. Age group results were slightly more varying: the 3-18 year
  olds remain having the highest epidemic potential, but from
  $\varepsilon>35$ days on, the oldest age group comes in second. The
  sensitivity analysis also showed, that there is too little
  information to estimate $\varepsilon$ from the IMD data -- we are
  thus forced to fix the hyperparameter at a biological plausible
  value.

\section{Discussion}
\label{sec:discussion}

We presented a comprehensive framework for modelling, inference and
simulation for infectious disease occurrence data. In the case of IMD,
the infected individual is effectively removed from the transmission
network once the disease becomes manifest.  Secondary cases are
thought to acquire the infective strain either from the case during
incubation or from asymptomatic carriers close to the case. Although
marks attached to the case can naturally not account for the latter
mode of transmission, they represent a valid proxy for the
transmission network of the case when analysing surveillance data,
which typically lack information regarding carriage.

Despite use of disease surveillance data, we were able to quantify
differences in IMD transmission dynamics based on age and finetype.
That the modelling requires an epidemic component is of epidemiologic
interest in its own, as this shows that IMD incidence goes beyond
sporadic occurrences. To our knowledge, our analysis is the first
report of finetype-specific differences in spread tendencies. Contrary
to previous analyses we were not able to find a significant connection between IMD
and concurrent number of influenza cases. The spatial spread appeared
to happen at a rather small scale -- a scale which the usual
district resolution data collected as part of the German Infection
Protection Act does not allow to analyse. Thus our work is also a
contribution to the controversy between patient privacy and the need
for high-resolution data to gain new epidemiological insights. One important
question in this debate is how good a proxy the patient's residence is for
his general whereabouts.
%It will be interesting to see, if an
%operationalisation of social contact network will become possible in
%the future.

Even though our CIF modelling is similar in form to the proposal in
\citet{Hoehle2009}, the \textit{continuous space} of the IMD
application makes epidemic modelling conceptually different.
The classical SIR
model framework does not apply in this situation, because events do
not originate from a predefined population and individuals can not be
partitioned into model compartments anymore. Thus, including
population density becomes important and one needs to distinguish
between covariate information of events and covariates of the
surrounding environment within which the process occurs. Furthermore,
likelihood inference is complicated by requiring an additional
integration over space for complex polygons. Finally, the now proposed
space-time interaction functions are completely general in form and
thus provide an advantage over the previous linear basis decomposition
and resulting parameter constraints.

An issue currently not dealt with in our estimation are edge effects,
i.e.\ data are only available for Germany, but infections occur
outside the observation window. For example, \citet{Elias2010}
investigate the contribution of cross-border spread to increased
incidence of IMD in the German region of Aachen neighbouring the
Netherlands. A cross-border effect is indeed detected by our
  simulation in Figure~\ref{fig:incidence+outliers} where the Aachen
  region has higher observed incidences than can be explained by our
  model. Hence, the actual disease clusters are wider than observed
in Germany, which potentially causes underestimation of the epidemic
weight. Edge correction for inference in spatio-temporal point
processes is, however, still an open methodological issue.

An additional strength of the proposed modelling is that it offers a
parametric framework for conducting prospective change-point analysis
in spatio-temporal point processes typical in disease surveillance:
Within the framework of stochastic process control one could e.g.\ use
likelihood ratio detectors to monitor the time point where inclusion
of an epidemic component is necessary to describe the observed
data. This would correspond in idea to the time series setting
investigated in \citet{hoehle_paul2008} or the homogeneous
spatio-temporal Poisson process setting of
\citet{assuncao_correa2009}.

The presented methods for inference and simulation of
\texttt{twinstim} models are available as part of the
\texttt{R} package \texttt{surveillance} \citep{R:surveillance,hoehle2007}
available from the Comprehensive \texttt{R} Archive Network.

%-SOME ADDITIONAL INFORMATION-------------------------------------

\section{Supplementary Materials}

The Web Animation referenced in Section~\ref{sec:intro} and
the Web Appendices referenced in Sections~\ref{sec:inference} and~\ref{sec:simulation}
are available under the Paper Information link at the \textit{Biometrics} website
\url{http://www.biometrics.tibs.org/}.

\backmatter

\section*{Acknowledgements}

We thank Ludwig Fahrmeir for providing helpful suggestions and
comments. Financial support was provided by the Munich Center of
Health Sciences.  Ulrich Vogel is thanked for his efforts in ensuring
the generation of high quality IMD surveillance data and helpful
discussions. Matthias Frosch is acknowledged for continuous support.
We thank the co-editor Thomas Louis, an anonymous associate editor
and two anonymous referees for their useful comments that improved
the presentation of the article.

% \section*{Conflict of Interests Statement}
%
% The authors have declared no conflict of interest.

%-BIBLIOGRAPHY----------------------------------------------------

%\renewcommand{\bibliographytypesize}{\small}
%\renewcommand{\BCBT}{} % apacite: no comma before "&" in author list
%\renewcommand{\BCBL}{} % apacite: no comma before "&" in author list
\bibliographystyle{biom}   %apalike, apalike2, elsart-harv, kluwer, chicago, munich
\begin{small}
\bibliography{bibliography}
\end{small}

%-APPENDICES------------------------------------------------------
%\appendix

\label{lastpage}

%\clearpage
%\pagestyle{empty}  % pages of figures and tables should not be numbered

\end{document}